\documentclass[a4paper,11pt]{article}
\usepackage{jheppub}
\usepackage{eurosym}
\usepackage{amssymb}
\usepackage{amsfonts,cite}
\usepackage{amsmath}
\usepackage{graphicx}
\usepackage{multirow}
\usepackage{xcolor}
\usepackage{epsfig}
\usepackage{fancyhdr}

\newbox\mybox

\newcommand\fverb{\setbox\mybox=\hbox\bgroup\verb}
\newcommand\fverbdo{\egroup\medskip\noindent\fbox{\unhbox\mybox}\ }
\newcommand\fverbit{\egroup\item[\fbox{\unhbox\mybox}]}
\abstract{We present a detailed analysis of the sixth-order Pais-Uhlenbeck oscillator and construct three-dimensional ghost-free representations through a Tri-Hamiltonian framework. We identify a six-dimensional Abelian Lie algebra of the PU model's dynamical flow and derive a hierarchy of conserved Hamiltonians governed by multiple compatible Poisson structures. These structures enable the realisation of a complete Tri-Hamiltonian formulation that generates identical dynamical flows. Positive-definite Hamiltonians are constructed, and their relation to the full Tri-Hamiltonian hierarchy is analysed. Furthermore, we develop a mapping between the PU model and a class of three-dimensional coupled second-order systems, revealing explicit conditions for ghost-free equivalence. We also explore the consequences of introducing interaction terms, showing that the multi-Hamiltonian structure is generally lost in such cases.}    

\title{\bfseries Three-dimensional ghost-free representations of the Pais-Uhlenbeck model  from Tri-Hamiltonians}

\author[a]{Alexander Felski,}
\author[b]{Andreas Fring,}
\author[b]{Bethan Turner}

\affiliation[a]{Max Planck Institute for the Science of Light, Staudtstraße 2,91058 Erlangen, Germany}
\affiliation[b]{Department of Mathematics, City St George's, University of London, Northampton Square, \\ London EC1V 0HB, UK}

\emailAdd{alexander.felski@mpl.mpg.de}
\emailAdd{a.fring@city.ac.uk}
\emailAdd{bethan.turner.2@city.ac.uk}

\begin{document}
	\maketitle
	
	\pagestyle{fancy}
	\fancyhead{} 
	\fancyhead[LE,RO]{\small\itshape  3D ghost-free representations of the Pais-Uhlenbeck model from Tri-Hamiltonians} 
	
	\renewcommand{\headrulewidth}{0.4pt}

\section{Introduction}	

We extend our prior analysis \cite{FFT} of the Pais-Uhlenbeck (PU) model \cite{pais1950field} in which we  proposed an approach that leads to a  unified framework for interpreting and stabilising higher time-derivative dynamics through a Lie symmetry analysis and the utilisation of a Bi-Hamiltonian structure. The scheme was applied to the fourth-order PU model. Here we extend the application to its sixth order variant that requires the use of a Tri-Hamiltonian structure, see e.g., \cite{oevel1990abstract,fordy1991tri,olver1996tri}. The sixth-order PU oscillator is of particular interest as it naturally accommodates three degrees of freedom, providing an ideal framework for studying three-dimensional representations of higher-derivative systems.

Our procedure consists of the following steps: i) Identify the Multi-Hamiltonian structure for the dynamical flow of the higher time-derivative equation of motion. ii) Identify the autonomous Lie symmetries for the vector equation of the dynamical flow. iii) Construct new dynamical flows from linear combinations of the Poisson bracket tensors and the multi-Hamiltonians. iv) Use the Lie symmetries to identify the constraints that preserve the original flow. v) Select all relevant solutions by imposing positive-definiteness on the Hamiltonian. vi) Map the higher time-derivative theory (HTDT) to coupled lower dimensional theories and identify the viable models in those higher dimensional spaces.

The quantisation of the positive-definite versions with the same dynamical flow as the Pais-Uhlenbeck system will lead to a consistent theory in which the usual deficiencies of HTDT, i.e. possessing unphysical ghost states that are non-normalisable or lead to unbounded spectra from below, are absent.  

Alternative proposals to address the ghost-state problem have been made in the past by several authors. One may for instance simply accept the presence of these states and argue that the effect they cause is not very severe and in particular that unitarity is restored at the low energies that now occur in the universe \cite{Hawking}. In an approach that assembles combination of Noether charges constructed from Lie symmetries positive Hamiltonians were constructed in \cite{nucci1,nucci2,nucci3}. However, as we have shown in \cite{FFT} those combinations have dynamical flows that are different from the ones in the PU oscillators and therefore describe different physical systems. Another approach consists of imposing constraints that ensure the disappearance of the Ostrogradsky instabilities \cite{ghostconst,motohashi1,motohashi2,motohashi3}. The precise physical nature and justification of these type of constraints still needs clarification. Further proposals include the use of Dirac–Pauli quantisation scheme with an indefinite metric \cite{salvio16quant}, the introduction of so-called fakeons \cite{fakeons}, i.e.  additional degrees of freedom that do not belong to the physical Hilbert space, the non-unitary mapping of the Hermitian Hamiltonian to another Hermitian Hamiltonian in which the non-normalisable sections of the theory become normalisable \cite{fring2025ghost} or the complex extensions of the models to a ${\cal PT}$-symmetric non-Hermitian system \cite{bender2008no,mosta2011imag,raidal2017quantisation}.

Our paper is organised as follows: In section 2 we begin by analysing the symmetries of the sixth-order PU model and deriving its corresponding Hamiltonian via Ostrogradsky’s method. We then identify six Lie symmetries, three of which generate new Hamiltonians in involution, enabling a Tri-Hamiltonian description of the system. In sections 3 we systematically apply these symmetries, deriving an entire hierarchy of ghost-free Hamiltonians and demonstrate how these can be embedded into a positive-definite formulation. In section 4 we explore mappings to three-dimensional coupled second-order systems, examining conditions under which these alternative representations in terms of standard canonical variables remain ghost-free. Finally, in section 5 we analyse the impact of adding interaction terms to the model, highlighting how these terms affect the underlying Hamiltonian structure. Our conclusions are stated in section 6.

\section{The sixth order PU model and its Lie symmetries}

The sixth order version of the PU model \cite{pais1950field} is an especially attractive version of the series, as it relates to a model in three space dimensions. Its defining Lagrangian density involves time-derivatives of the coordinate $q$ up to order three
\begin{equation}
	{\cal L}_{\text{PU}}(q,\dot{q} , \ddot{q },  \dddot{q })=\frac{1}{2}  \dddot{q}^2  -\frac{\alpha}{2}  \ddot{q}^2 + \frac{\beta}{2} \dot{q}^2 - \frac{\gamma}{2} q^2 , \qquad \alpha,\beta,\gamma \in \mathbb{R} .  \label{LPU}
\end{equation}
We denote partial derivatives either by overdots or explicit subscripts. The corresponding Euler-Lagrange equation resulting from $	{\cal L}_{\text{PU}}(q,\dot{q} , \ddot{q}, \dddot{q })$ is the sixth order PU oscillator equation  
\begin{equation}
	\frac{d^3}{dt^3}   \frac{	\delta {\cal L}}{\delta \dddot{q}}-	\frac{d^2}{dt^2}   \frac{	\delta {\cal L}}{\delta \ddot{q}} + \frac{d}{dt} \frac{\delta  {\cal L} }{\delta \dot{q}} - \frac{\delta  {\cal L}}{\delta q}  =0, \qquad \Rightarrow \qquad   q_{6t}  + \alpha q_{4t} + \beta q_{2t}  + \gamma q=0. \label{equm1}
\end{equation}
When more convenient we frequently use the parametrisation  
\begin{equation}
	\alpha = \omega_1^2 + \omega_2^2+ \omega_3^2, \quad
	\beta = \omega_1^2  \omega_2^2  + \omega_1^2  \omega_3^2  + \omega_2^2  \omega_3^2, \quad 
	\gamma=  \omega_1^2  \omega_2^2 \omega_3^2 . \label{param}
\end{equation}
This is the case in the general real solutions that are purely oscillatory 
\begin{equation}
	q(t) = a_1 \sin( \omega _1 t) + a_2 \cos( \omega _1 t) + b_1 \sin( \omega _2 t) + b_2 \cos( \omega _2 t)
	+ c_1 \sin( \omega _3 t) + c_2 \cos( \omega _3 t) \label{solndeg}
\end{equation}
in the nondegenerate case $\omega _1  \neq \omega _2 \neq \omega _3 $, and oscillatory as well as asymptotically divergent
in the partially degenerate case, when e.g. $\omega _1 =\omega _2 =: \omega $
\begin{equation}
	q(t) = a_1 \sin( \omega  t) + a_2 \cos( \omega  t) + b_1  t \sin( \omega  t) + b_2 t \cos( \omega  t)
	+ c_1 \sin( \omega _3 t) + c_2 \cos( \omega _3 t) , \label{soldegpart}
\end{equation}
and the fully degenerate case when $\omega _1 =\omega _2=\omega _3 =: \omega $
\begin{equation}
	q(t) =  c_1 \sin ( \omega t ) + c_2 \cos ( \omega t ) + c_3 t \sin ( \omega t ) + c_4 t \cos ( \omega t ) 
	+ c_5 t^2 \sin ( \omega t ) + c_6 t^2 \cos ( \omega t ) .
\end{equation}

Following Ostrogradsky's approach \cite{ostrogradsky1850memoire} to carry out a Legendre transformation we derive the corresponding Hamiltonian. According to  \cite{ostrogradsky1850memoire}, the canonical coordinates $q_i$ and momenta $\pi_i$, $i=1,2,3$ are identified as  
\begin{eqnarray}
	q_1 &=& q,    \quad   q_2 = \dot{q},   \quad    q_3 = \ddot{q},   \label{cano1}  \\
    \pi_1 &=&  \frac{\partial {\cal L} }{\partial \dot{q}} - \frac{d}{dt}  \frac{\partial {\cal L}}{\partial \ddot{q}}  
    + \frac{d^2}{dt^2}  \frac{\partial {\cal L}}{\partial \dddot{q}}
    =  \beta \dot{q}   + \alpha \dddot{q} + q_{5t},   \quad \pi_2 = \frac{\partial {\cal L}}{\partial \ddot{q}}  - \frac{d}{dt}  \frac{\partial {\cal L}}{\partial \dddot{q}}   = - \alpha \ddot{q} - q_{4t} ,  \quad  \\
	    	\pi_3 &=& \frac{\partial {\cal L}}{\partial \dddot{q}}  = \dddot{q} .   \label{cano3}
\end{eqnarray}	
The Hamiltonian then results to 
\begin{equation}
	{\cal H}_{\text{PU}}(q_1,q_2,q_3,\pi_1,\pi_2,\pi_3)= \pi_1 q_2 + \pi_2 q_3 + \frac{1}{2}  \pi_3^2 
	+\frac{\alpha}{2} q_3^2 - \frac{\beta}{2} q_2^2 + \frac{\gamma}{2} q_1^2  .
\end{equation}
One can easily verify that Hamilton's relations 
\begin{equation}
	\dot{q}_i = \frac{{\cal H}_{\text{PU}}}{\partial \pi_i   } = \left\{   q_i ,  {\cal H}_{\text{PU}}   \right\}_c,   \quad
	\dot{\pi}_i = -\frac{{\cal H}_{\text{PU}}}{\partial q_i   } = \left\{   \pi_i ,  {\cal H}_{\text{PU}}   \right\}_c, \quad
	\left\{   q_i , \pi_j    \right\}_c= \delta_{ij} ,   \quad i,j = 1,2,3,
\end{equation}
are obtained from this Hamiltonian with the standard canonical Poisson bracket relations. Converting to the original PU variables $ \vec{q} = \left(q,\dot{q} , \ddot{q },  \dddot{q },q_{4t},q_{5t}  \right)$  by means of (\ref{cano1})-(\ref{cano3}) this Hamiltonian acquires the form
\begin{equation}
	{\cal H}_{1}( \vec{q})=  \frac{\dddot{q}^2}{2} -\frac{\alpha  }{2} \ddot{q}^2
	+\frac{\beta 	}{2} \dot{q}^2 +	\frac{\gamma  }{2} q^2 +\alpha  \dot{q} \dddot{q}-q_{4 t} \ddot{q}+\dot{q} q_{5 t}, \label{Ham1}
\end{equation}
whereby the only non-vanishing Poisson brackets in these variables are
\begin{equation}
\left\{ q ,q_{5t}      \right\}=-\left\{ \dot{q} ,q_{4t}      \right\}=
\left\{ \ddot{q} , \dddot{q}      \right\}=1, \quad \left\{ \ddot{q} ,q_{5t}      \right\}=
-\left\{ \dddot{q} ,q_{4t}      \right\}=-\alpha, \quad  \left\{ q_{4t} ,q_{5t}      \right\}=\alpha^2 - \beta .     \label{Pbrac}
\end{equation}

\subsection{Lie symmetries of the sixth-order PU oscillator equation}
The dynamical equations of the PU oscillator can be cast into the form
\begin{equation}
	\frac{d\vec{q}}{dt} = \vec{V}(\vec{q}),   \quad  V= \sum_{i=1}^6 v_i \partial_{q_i}   = \dot{q} \partial_q +  \ddot{q} \partial_{\dot{q}} 
	+  \dddot{q} \partial_{\ddot{q}} +  q_{4t} \partial_{\dddot{q}} +  q_{5t} \partial_{  q_{4t} }
	- \left( \alpha q_{4t} + \beta \ddot{q}  + \gamma q \right) \partial_{ q_{5t} },  \label{1flow}
\end{equation}
where $\vec{v} =  ( \dot{q} ,  \ddot{q}
, \dddot{q} ,  q_{4t}  , q_{5t},
-  \alpha q_{4t} - \beta \ddot{q}  - \gamma q ) $ and the first-order linear differential operator $V$ acts like a vector field when interpreted as a derivation on $C^{\infty}(\mathbb{R}^6)$.

Next we identify the Lie symmetries of the dynamical system in (\ref{1flow}) in the usual way, see e.g. \cite{hydonsymm}. Infinitesimally transforming $q_i \rightarrow \tilde{q}_i = q_i + \varepsilon \xi_i$, $\varepsilon \ll 1$, the component version of (\ref{1flow}) transforms to
\begin{equation}
     \frac{d \tilde{q}_i}{dt} = V_i(\tilde{q}) - \delta V_i(\tilde{q}), \qquad \text{with} \,\,\,\,  \delta V_i(\tilde{q}) = \varepsilon \left[X,V\right]_i(\tilde{q}) = \varepsilon  \sum_{j=1}^6   \left(  \xi_j \frac{\partial v_i}{ \partial q_j} - v_j \frac{\partial \xi_i}{ \partial q_j} \right),
\end{equation}
where the Lie algebraic generators are
\begin{equation}
   X = \sum_{i=1}^6 \xi_i \partial_{q_i}  .
\end{equation}
Using a generic Ansatz for the vector field $\xi_i$ as being linear in the coordinates $q_i$, we find six linearly independent solutions for the Lie bracket to vanish  
\begin{eqnarray}
	X_1 &=&  \dot{q} \partial_q +  \ddot{q} \partial_{\dot{q}} 
	+  \dddot{q} \partial_{\ddot{q}} +  q_{4t} \partial_{\dddot{q}} +  q_{5t} \partial_{  q_{4t} }
	- \left( \alpha q_{4t} + \beta \ddot{q}  + \gamma q \right) \partial_{ q_{5t} }, \label{Lie1} \\
	X_2 &=&  \left(\alpha  \dot{q}+\dddot{q}\right)  \partial_q
	+ \left(\alpha  \ddot{q}+q_{4t}\right)   \partial_{\dot{q}}
	+ \left(\alpha  \dddot{q}+q_{5 t}\right)  \partial_{\ddot{q}}
	- \left(\beta 	\ddot{q}+\gamma  q\right)    \partial_{\dddot{q}}      \label{Lie4}  \\
	&& \qquad   \qquad    \qquad    \qquad    \qquad   \qquad    \qquad   \qquad     \qquad	- \left(\beta  \dddot{q}+\gamma  \dot{q}\right)     \partial_{q_{4t}}
	- \left(\gamma  \ddot{q}+\beta  q_{4 t}\right)   \partial_{q_{5t}} ,  \notag \\
	X_3 &=&  \left(\alpha  \dddot{q}+\beta  \dot{q}+q_{5 t}\right)  \partial_q   -\gamma  \left(
	q  \partial_{\dot{q}} + \dot{q}  \partial_{\ddot{q}}+ \ddot{q}  \partial_{\dddot{q}} + \dddot{q}  \partial_{q_{4t}} +\ q_{4 t}  \partial_{q_{5t}}  \right), \\
		X_4 &=&   \frac{1}{2} \left[  q \partial_q +  \dot{q} \partial_{\dot{q}} +
		  \ddot{q} \partial_{\ddot{q}} + \dddot{q}   \partial_{ \dddot{q} }   + q_{4t}   \partial_{ q_{4t}  }  + q_{5t}   \partial_{ q_{5t}  }      \right],      \label{Lie2}        \\	  
		  	X_5 &=& \frac{1}{2}   \left[ \left(\alpha  \ddot{q}+\beta  q+q_{4 t}\right)  \partial_q
		  + \left(\alpha \dddot{q}+\beta  \dot{q}+q_{5 t}\right)    \partial_{\dot{q}}
		  - \gamma  \left( q   \partial_{\ddot{q}}
		  +  \dot{q}    \partial_{\dddot{q}}
		  +  \ddot{q}    \partial_{q_{4t}}
		  +   \dddot{q}   \partial_{q_{5t}}  \right)  \right], \qquad \,\, \\
		X_6 &=&  \frac{1}{2} \left\{ \left( \alpha  \beta  \ddot{q}-\alpha  \gamma  q+\beta ^2 q-\gamma  \ddot{q}+\beta  q_{4 t}     \right) \partial_q
		+ \left(  \alpha  \beta  \dddot{q}-\alpha  \gamma  \dot{q}+\beta ^2 \dot{q}-\gamma  \dddot{q}+\beta  q_{5 t}     \right)  \partial_{\dot{q}}
	  \right.   \label{Lie6}   \\
		&&   \left. \qquad \qquad 	- \gamma \left(   \beta q +  \alpha \ddot{q} + q_{4t}         \right)       \partial_{\ddot{q}} 
		- \gamma \left(   \beta \dot{q} +  \alpha \dddot{q} + q_{5t}         \right)       \partial_\dddot{q}  
		+ \gamma^2 q  \partial_{q_{4t}}
		+ \gamma^2 \dot{q}  \partial_{q_{5t}}   \right\}.       \notag     
\end{eqnarray}
The algebra formed by these generators is Abelian, i.e $[X_i,X_j]=0$, for $i,j=1,\ldots,6$. 
\subsection{Tri-Hamiltonians}
Next we consider the action of these Lie derivatives on the Hamiltonian ${\cal H}_{1}$. We compute that 
$X_1$, $X_2$ and $X_3$ are symmetries of the PU Hamiltonian, that is $ X_1( {\cal H}_{1})=X_2( {\cal H}_{1}) =X_3( {\cal H}_{1}) =0$. This is obvious for $X_1$, as it is simply the dynamical vector field $V.$  $X_4$  is the Euler symmetry operator that maps ${\cal H}_{1}$ to itself, whereas $X_5$ and $X_6$ define new operators, $ X_5( {\cal H}_{1}) =: {\cal H}_{2}$, $ X_6( {\cal H}_{1}) =: {\cal H}_{3}$  that we can interpret as new Hamiltonians generating the same flow with the appropriate Poisson bracket tensors. Concretely, we obtain
\begin{eqnarray}
	{\cal H}_2\left( \vec{q} \right) &=& 
	\frac{1}{2} \left(\alpha  \dddot{q}+\beta  \dot{q}+q_{5 t}\right)^2 +\frac{\gamma}{2} \left[   \beta  q^2+2 \alpha  q \ddot{q}-\alpha  \dot{q}^2+2 q q_{4 t}+\ddot{q}^2-2 \dot{q} \dddot{q}     \right]  	,  \label{Ham2} \\
	{\cal H}_3\left( \vec{q} \right) &=& \frac{1}{2} \beta  \left(\alpha  \dddot{q}+\beta  \dot{q}+q_{5 t}\right)^2  + \frac{\gamma}{2}   \left\{   \dot{q}^2 (\gamma -2 \alpha  \beta )+\left(\alpha  \ddot{q}+\beta  q\right)^2- \gamma  q \left(\alpha  q+2 \ddot{q}\right) \right. \quad \\
	&&\left. 
	-2 \dot{q}
	\left[    \left(\alpha ^2+\beta \right) \dddot{q}+\alpha  q_{5 t}\right]+2 q_{4 t} \left(\alpha  \ddot{q}+\beta  q\right)-2 \dddot{q} \left(\alpha 
	\dddot{q}+q_{5 t}\right)+q_{4 t}^2\right\}   . \label{Ham3}   \notag
\end{eqnarray}
Casting the Poisson bracket relations (\ref{Pbrac}) into a matrix gives
\begin{equation}
	J_1  = \left(
	\begin{array}{cccccc}
		0 & 0 & 0 & 0 & 0 & 1 \\
		0 & 0 & 0 & 0 & -1 & 0 \\
		0 & 0 & 0 & 1 & 0 & -\alpha  \\
		0 & 0 & -1 & 0 & \alpha  & 0 \\
		0 & 1 & 0 & -\alpha  & 0 & \alpha ^2-\beta  \\
		-1 & 0 & \alpha  & 0 & \beta -\alpha ^2 & 0 \\
	\end{array}
	\right),
\end{equation}
we can solve the flow equations 
\begin{equation}
	\vec{v} = (\dot{q},  \ddot{q},
	  \dddot{q} ,  q_{4t} , q_{5t} ,
	- \left( \alpha q_{4t} + \beta \ddot{q}  + \gamma q \right))=
	J_1 \nabla{\cal H}_1 = J_2 \nabla{\cal H}_2  = J_3 \nabla{\cal H}_3   ,    \label{flow1}
\end{equation}
for the Poisson tensors $J_2$ and $J_3$. We find
\begin{equation}
	J_2=  \frac{1}{\gamma} \left(
	\begin{array}{cccccc}
		0 & 0 & 0 & -1 & 0 & \alpha  \\
		0 & 0 & 1 & 0 & -\alpha  & 0 \\
		0 & -1 & 0 & \alpha  & 0 & \beta -\alpha ^2 \\
		1 & 0 & -\alpha  & 0 & \alpha ^2-\beta  & 0 \\
		0 & \alpha  & 0 & \beta -\alpha ^2 & 0 & \delta_1   \\
		-\alpha  & 0 & \alpha ^2-\beta  & 0 & -\delta_1  & 0 \\
	\end{array}
	\right),
\end{equation}
and
\begin{equation}
	J_3=  \frac{1}{\gamma^2} \left(
	\begin{array}{cccccc}
		0 & 1 & 0 & -\alpha  & 0 & \alpha ^2-\beta  \\
		-1 & 0 & \alpha  & 0 & \beta -\alpha ^2 & 0 \\
		0 & -\alpha  & 0 & \alpha ^2-\beta  & 0 & -\delta_1   \\
		\alpha  & 0 & \beta -\alpha ^2 & 0 & \delta_1  & 0 \\
		0 & \alpha ^2-\beta  & 0 & -\delta_1   & 0 &    \delta_2 \\
		\beta -\alpha ^2 & 0 &\delta_1  & 0 & -  \delta_2 & 0 \\
	\end{array}
	\right) ,
\end{equation}  
where we abbreviated $\delta_1 = \alpha ^3-2 \alpha  \beta +\gamma  $ and $\delta_2 = \alpha ^4-3 \alpha ^2
\beta +2 \alpha  \gamma +\beta ^2  $. 

In reverse, we could have constructed all anti-symmetric tensors with vanishing Lie derivatives
\begin{align}
	L_{V}\left( J \right) = V^{k}  \partial_k J^{i,j} - \partial_k V^i  J^{k,j} - J^{i,k} \partial_k V^j = 0 , \label{Lieder}
\end{align}
establishing that $V$ is a Poisson vector field for the Poisson tensors $J_{1,2,3}$. Since $J_{1,2,3}$ are constant and invertible, $\det J_1 =1$,  $\det J_2 =\gamma^{-4}$, $\det J_3 =\gamma^{-8}$, this implies that $V$ is also a Hamiltonian vector field on simply connected domains \cite{spivak2018ca,dam2006}, so that  $J^{-1} \vec{v} = \nabla {\cal H}$ can be solved for ${\cal H}$.

Having obtained three solutions, we constructed a Tri-Hamiltonian structure, i.e. three independent Hamiltonians with appropriate Poisson brackets that reproduce exactly the same flow.

Let us next see how the Lie symmetries act on the Hamiltonians  ${\cal H}_{2}$ and  ${\cal H}_{3}$. We find
\begin{equation}
	X_1({\cal H}_{i}) = X_2({\cal H}_{i})=X_3({\cal H}_{i})=0, \,\, X_4({\cal H}_{i}) =  {\cal H}_{i}, 
	 \,\, X_5({\cal H}_{i}) =  {\cal H}_{i+1}, \,\, X_6({\cal H}_{i}) =  {\cal H}_{i+2} ,  \label{actLie}
\end{equation}
for $i \in \mathbb{N}$, defining in this way the new Hamiltonians ${\cal H}_{i}$ for $i>3$. Similarly to Bi-Hamiltonian systems, see e.g. \cite{Das,Magri}, these relations can be iterated as
\begin{equation}
J_3 \nabla 	{\cal H}_{n+2} =	J_2 \nabla 	{\cal H}_{n+1} = J_1 \nabla 	{\cal H}_n, \qquad n \in \mathbb{N},   \label{recn}
\end{equation}
thus generating new Tri-Hamiltonian flows. By construction all higher Hamiltonians are conserved $d{\cal H}_{i}/dt=0$ and in involution 
$\{ {\cal H}_{i} , {\cal H}_{j}    \}_1=\{ {\cal H}_{i} , {\cal H}_{j}    \}_2 =\{ {\cal H}_{i} , {\cal H}_{j}    \}_3 =0$. 

We can use (\ref{actLie}) to construct the entire hierarchy of Hamiltonians. Defining the vector $\vec{{\cal H}} = ({\cal H}_1,{\cal H}_2,{\cal H}_3)$, we cast the consecutive action $X_5$ on the Hamiltonians in matrix form as
\begin{equation}
	   X_5^k (\vec{{\cal H}}) = M^k (\vec{{\cal H}}), \qquad   \text{with} \,\, M=\left(
	   \begin{array}{ccc}
	   	0 & 1 & 0 \\
	   	0 & 0 & 1 \\
	   	\gamma ^2 & -\alpha  \gamma  & \beta  \\
	   \end{array}
	   \right) .
\end{equation}
Diagonalising $M$, the $k^{ \text{th}} $ power of $M$ is easily computed to
\begin{equation}
	M^k = U D^k U^{-1}, \qquad   U = \left(
	\begin{array}{ccc}
		\frac{1}{\omega _1^4 \omega _2^4} & \frac{1}{\omega _1^4 \omega _3^4} & \frac{1}{\omega _2^4
			\omega _3^4} \\
		\frac{1}{\omega _1^2 \omega _2^2} & \frac{1}{\omega _1^2 \omega _3^2} & \frac{1}{\omega _2^2
			\omega _3^2} \\
		1 & 1 & 1 \\
	\end{array}
	\right), \qquad
	D = \left(
	\begin{array}{ccc}
		\omega _1^2 \omega _2^2 & 0 & 0 \\
		0 & \omega _1^2 \omega _3^2 & 0 \\
		0 & 0 & \omega _2^2 \omega _3^2 \\
	\end{array}
	\right) .
\end{equation}
Acting on $\vec{{\cal H}}$ we obtain the entire hierarchy of Hamiltonians in closed form
\begin{eqnarray}
{\cal H}_n &=& \left[ \frac{\omega _3^4 \left(\omega _1^2 \omega _2^2\right)^{n+1}}{\left(\omega _1^2-\omega _3^2\right)
	\left(\omega _2^2-\omega _3^2\right)}+\frac{\omega _1^4 \left(\omega _2^2 \omega
	_3^2\right)^{n+1}}{\left(\omega _1^2-\omega _2^2\right) \left(\omega _1^2-\omega
	_3^2\right)}+\frac{\omega _2^4 \left(\omega _1^2 \omega _3^2\right)^{n+1}}{\left(\omega
	_2^2-\omega _1^2\right) \left(\omega _2^2-\omega _3^2\right)}   \right]  {\cal H}_1  \qquad \label{Hnlong}  \\
	&&  \!\! \!\! \!\! \!\! \!\!   + \left[  \frac{\omega _2^2 \left(\omega _1^2+\omega _3^2\right) \left(\omega _1^2 \omega
		_3^2\right)^{n}}{\left(\omega _1^2-\omega _2^2\right) \left(\omega _2^2-\omega
		_3^2\right)}
	-\frac{ \omega _3^2 \left(\omega _1^2+\omega _2^2\right)  \left(\omega _1^2 \omega
		_2^2\right)^{n}}{\left(\omega _1^2-\omega _3^2\right) \left(\omega _2^2-\omega
		_3^2\right)}
		-\frac{\omega _1^2 \left(\omega _2^2+\omega _3^2\right) \left(\omega _2^2 \omega
		_3^2\right)^{n}}{\left(\omega _1^2-\omega _2^2\right) \left(\omega _1^2-\omega
		_3^2\right)}  \right]    {\cal H}_2  \notag \\
	&&   \!\! \!\! \!\! \!\! \!\!    +\left[   \frac{\left(\omega _1^2 \omega _2^2\right)^{n}}{ \left(\omega _1^2-\omega
		_3^2\right) \left(\omega _2^2-\omega _3^2\right)}+\frac{\left( \omega _2^2 \omega
		_3^2\right)^{n}}{ \left(\omega _1^2-\omega _2^2\right)  \left(\omega
		_1^2-\omega _3^2\right)}-\frac{\left( \omega _1^2 \omega _3^2\right)^{n}}{
		\left(\omega _1^2-\omega _2^2\right)\left(\omega _2^2-\omega _3^2\right)}            \right]   {\cal H}_3 .  \notag
\end{eqnarray}  
Note that the entire hierarchy can be expressed in linear combinations of ${\cal H}_1$, ${\cal H}_2$ and ${\cal H}_3$. 

Next we compute the flow resulting from linear combinations of the Poisson tensors and Hamiltonians. Defining
\begin{equation}
	\bar{J} :=  c_1 J_1 +  c_2 J_2 + c_3 J_3, \qquad \text{and} \qquad    \bar{{\cal H}} :=   c_4 {\cal H}_{1} +  c_5 {\cal H}_{2}+  c_6{\cal H}_{3}. \label{barJH}
\end{equation}
we compute 
\begin{eqnarray}
		\bar{J} \nabla \bar{ {\cal H} } &=& 	\left[ \frac{ \alpha ^2-\beta }{\gamma ^2} c_3 c_4 +\frac{\alpha }{\gamma }   \left(c_2 c_4+c_3
		c_5\right)+c_1 c_4+c_2 c_5+c_3 c_6 \right] 
	X_1(\vec{q})   \\ 
	&& -
	\left[  \frac{\alpha  }{\gamma ^2}c_3 c_4+\frac{c_2 c_4+c_3 c_5}{\gamma }+\gamma  c_1 c_6    \right]  X_2(\vec{q}) 	+
	\left[ \frac{c_3 c_4}{\gamma ^2}+  \beta  c_6  c_1 +c_5  c_1 +c_2 c_6    \right]  X_3(\vec{q}).   \notag
\end{eqnarray}
Thus, setting the coefficient factor in the first term to 1 and in the second and third to 0, we recover the standard flow
\begin{equation}
	\frac{d \vec{q}}{ dt} =  \bar{J} \nabla \bar{{\cal H}} = V(\vec{q}).  \,\,
	  \label{a1234}
\end{equation}
 We may solve these equations for given Poisson bracket tensor or given Hamiltonian. Opting for the latter we find the solution
 \begin{eqnarray}
 	c_1&=& \frac{-\alpha  \gamma  c_6 c_4-\gamma  c_5 c_6+c_4^2}{\left(c_6 \omega _1^4 \omega _2^4+c_5
 		\omega _1^2 \omega _2^2+c_4\right) \left(c_6 \omega _1^4 \omega _3^4+c_5 \omega _1^2 \omega
 		_3^2+c_4\right) \left(c_6 \omega _2^4 \omega _3^4+c_5 \omega _2^2 \omega _3^2+c_4\right)}	, \qquad  \label{c1sol}  \\
 		c_2&=& 	\frac{\alpha  c_4 c_5 \gamma +\beta  c_5 c_6 \gamma ^2+\beta  c_4 c_6 \gamma +c_5^2 \gamma
 			^2}{\left(c_6 \omega _1^4 \omega _2^4+c_5 \omega _1^2 \omega _2^2+c_4\right) \left(c_6 \omega
 			_1^4 \omega _3^4+c_5 \omega _1^2 \omega _3^2+c_4\right) \left(c_6 \omega _2^4 \omega _3^4+c_5
 			\omega _2^2 \omega _3^2+c_4\right)}, \\
 		c_3&=& 	\frac{-\beta  c_4 c_6 \gamma ^2+c_6^2 \gamma ^4-c_4 c_5 \gamma ^2}{\left(c_6 \omega _1^4 \omega
 			_2^4+c_5 \omega _1^2 \omega _2^2+c_4\right) \left(c_6 \omega _1^4 \omega _3^4+c_5 \omega _1^2
 			\omega _3^2+c_4\right) \left(c_6 \omega _2^4 \omega _3^4+c_5 \omega _2^2 \omega
 			_3^2+c_4\right)} .    \label{c3sol}
 \end{eqnarray}
We have now many options to choose our coefficient and can potentially construct combinations for $\bar{ {\cal H} } $ that are positive-definite.

\section{Positive-definite versions of the sixth-order PU model}

Just as for the treatment of the fourth order PU model \cite{FFT}, it is useful to define some positive-definite quantities
\begin{equation}
	H_{jk}= \vert  \epsilon_{ijk} \vert \left\{  
	\left[  q_{5 t}+   \dddot{q} \left(\omega _j^2+\omega
	_k^2\right)+\dot{q} \omega _j^2 \omega _k^2  \right]^2  + \omega _i^2 \left[ q_{4 t} +     \ddot{q} \left(\omega _j^2+\omega _k^2\right)+q \omega _j^2 \omega _k^2   \right]^2         \right\} ,  \label{posdef}
\end{equation}
with $\epsilon_{ijk} $ denoting the Levi-Civita tensor. These quantities transform under the Lie symmetries as 
\begin{eqnarray}
	X_1(H_{jk})&=& X_2(H_{jk})  = X_3(H_{jk}) =0, \quad \\
	X_4(H_{jk})&=&H_{jk}, \quad X_5(H_{jk})= \omega_j^2  \omega_k^2 H_{jk},  \quad 
	X_6(H_{jk})= \omega_j^4  \omega_k^4 H_{jk} .
\end{eqnarray}	
We can re-cast the entire sequence of Hamiltonians in a simpler closed form than (\ref{Hnlong}) as
\begin{equation}
	{\cal H}_{n}= \sum_{i,j,k=1}^3  \vert  \epsilon_{ijk} \vert         \frac{\omega_j^{2n-2} \omega_k^{2n-2} }{4 \left(\omega_i ^2 -\omega_j ^2\right)  \left(\omega_i ^2 -\omega_k^2\right)} H_{jk} , \quad \omega_i \neq \omega_j \neq  \omega_k .\label{Hnclosed}
\end{equation}
We notice that we cannot express any of the Hamiltonians ${\cal H}_{n}$ in terms of the positive quantities for the degenerate or partially degenerate theory. Moreover, we observe that 
while all $H_{jk}$ are positive-definite, we cannot achieve all its prefactors in the expansion of the ${\cal H}_{i}$ to be positive, so that none of the Hamiltonians ${\cal H}_{i}$ is fully positive-definite when expressed in terms of these quantities.  However, using combinations, a simple positive-definite Hamiltonian is for instance
\begin{equation}
	H = H_{12}+ H_{23}+ H_{13} = 2 \left[ (\alpha^2 - 2\beta)  {\cal H}_{1} + \left(3 - \frac{\alpha \beta}{\gamma}\right) {\cal H}_{2} +  \frac{\alpha}{\gamma}{\cal H}_{3} \right]  .
\end{equation} 
In a more systematic approach we use (\ref{Hnclosed}),
 and (\ref{barJH}) to re-express  $ \bar{{\cal H}}$ in terms of our positive-definite quantities (\ref{posdef}) and also (\ref{c1sol})-(\ref{c3sol}) to trade the constants $c_4, c_5, c_6$ for the $c_1, c_2, c_3$  
\begin{eqnarray}
	 \bar{{\cal H}}  &=&   \sum_{i,j,k=1}^3  \vert  \epsilon_{ijk} \vert  \frac{   \left( c_4+ c_5 \omega_j^2 \omega_k^2+ c_6 \omega_j^4 \omega_k^4          \right) }{4    \left( \omega_j^2 - \omega_i^2       \right)     \left( \omega_k^2 - \omega_i^2       \right)      }      H_{jk}  \label{e36}  \\ 
	  &=&  \sum_{i,j,k=1}^3  \vert  \epsilon_{ijk} \vert  \frac{\omega_j^4 \omega_k^4 }{4  \left( c_3 + c_2 \omega_j^2 \omega_k^2+ c_1 \omega_j^4 \omega_k^4          \right)  \left( \omega_j^2 - \omega_i^2       \right)     \left( \omega_k^2 - \omega_i^2       \right)      }      H_{jk}  .
\end{eqnarray}
Thus $ \bar{{\cal H}}  $ is positive-definite when all three prefactors of $H_{jk}$ are positive. There is no solution for this when any of the $c_1$, $c_2$ or $c_3$ are zero. Thus we need the entire Tri-Hamiltonian structure to achieve this. For a particular ordering, say $\omega_1> \omega_2 > \omega_ 3 >0$ this boils down to
\begin{eqnarray}
	c_3 + c_2 \omega_1^2 \omega_3^2 + c_1  \omega_1^4 \omega_3^4 >0 & \land&  \label{HB1} \\
	\left[  c_3 + c_2 \omega_2^2 \omega_3^2 + c_1  \omega_2^4 \omega_3^4 >0 \right. & \land& - c_1 \omega_2^2 \left( \omega_1^2 + \omega_3^2      \right)  < c_2 < -c_1 \omega_3^2 (\omega_1^2 + \omega_2^2) ,  \\
 \lor \,\,	c_3 + c_2 \omega_1^2 \omega_2^2 + c_1  \omega_1^4 \omega_2^4 >0 & \land& - c_1 \omega_1^2 \left( \omega_2^2 + \omega_3^2      \right)  < c_2 < -c_1 \omega_2^2 (\omega_1^2 + \omega_3^2)  \left. \right]  . \label{HB2}
\end{eqnarray} 
Using the version (\ref{HB1}) instead of (\ref{HB2}), the same relations hold with $c_1 \rightarrow c_6$, $c_2 \rightarrow c_5$ and $c_3 \rightarrow c_4$. In all of the positive-definite versions the ghost-state problem is expected to be absent.

\section{Three-dimensional first-order representations}	

Next we reformulate the theories from the previous discussion to a more conventional setting in terms of canonical variables. This can be achieved by exploring systematically some specific maps of the equations of motion resulting from the Lagrangian density ${\cal L}(q,\dot{q} , \ddot{q}, \dddot{q})$ in (\ref{LPU}) to those arising from a three-dimensional Lagrangian involving at most first-order time derivatives 
\begin{equation}
	{\cal L}(x,y,z,\dot{x}, \dot{y},\dot{z}) = \frac{a_x}{2} \dot{x}^2 +  \frac{a_y}{2} \dot{y}^2 +  \frac{a_z}{2} \dot{z}^2 -  \frac{b_x}{2} x^2 -  \frac{b_y}{2} y^2  -  \frac{b_z}{2} z^2  - g_1 x y  - g_2 x z - g_3 y z,   \label{genLxy}
\end{equation}
with $ a_i,b_i,g_j\in \mathbb{R} $ for $i=x,y,z$, $j=1,2,3$.
The transformations are assumed to be of the linear form
\begin{equation}
	x= \mu_0 q +  \mu_2 \ddot{q} +   \mu_4 q_{4t} , \quad
	y= \nu_0 q + \nu_2 \ddot{q} +   \nu_4 q_{4t}, \quad 
	z= \tau_0 q +  \tau_2 \ddot{q} +   \tau_4 q_{4t},    \label{pointt}
\end{equation}
with real constants $\mu_i, \nu_i, \tau_i$, $i=0,2,4$ that need to be determined. Without loss of generality we have already taken into account that terms with odd order time-derivatives are forced to vanish. Being three-dimensional, the Lagrangian density ${\cal L}(x,y,z,\dot{x}, \dot{y},\dot{z}) $ will give rise to three coupled second-order equations of motion
\begin{equation}
	a_x \ddot{x}  + b_x x + g_1 y + g_2 z =0, \quad 
	a_y \ddot{y}  + b_y y + g_1 x + g_3 z  =0, \quad 
	a_z \ddot{z}  + b_z z + g_2 x   + g_3 y =0 .  \label{equnmxy}
\end{equation}
Next we make contact with the PU sixth-order equation of motion in (\ref{equm1}). While it is in principle possible to follow the procedure in \cite{FFT} and classify all viable linear maps between the sixth-order equation (\ref{LPU}) and the three second-order equations (\ref{equnmxy}), the generic expressions become extremely cumbersome to present. We will therefore confine ourselves here to some simplified cases with specified constants. We distinguish between three different scenarios Ta when all three equation in (\ref{equnmxy}) transform individually into the PU-equation, Tb when two equations transform into the PU-equation with the remaining one vanishing trivially and Tc when only one equation transforms into the PU-equation with the remaining two vanishing trivially. We find the following solutions:

\noindent {\bf Ta1:}\\
\begin{equation}
	g_1=\frac{1}{2}\left( \rho_2 -1 +\alpha -\beta     \right) -\rho_1, \,\,\,\,\,
	g_2= \frac{1}{2} + \beta - \rho_2, \,\,\,\,\,
	g_3 = \frac{1}{2}\left( \rho_2 -1 +\alpha -\beta     \right) +\rho_1,
\end{equation}
\begin{align}
	\mu_0 & =  \frac{1}{2}\left( 1-\rho_2  -\alpha + 3\beta     \right) +\rho_1 ,  & \mu_2=0, \,\,  \qquad  &  \mu_4=1,    &    b_x = \frac{1}{2}\left( \rho_2  +\alpha -\beta     \right) +\rho_1 , \\
	\nu_0 & =   \rho_2  ,                                                                                         & \nu_2=1,  \,\,    \qquad   &  \nu_4 =1,   &    b_y =  \beta - \rho_2 ,  \,\, \,\, \quad   \qquad   \qquad \\
	\tau_0 & =   \frac{1}{2}\left( 1-\rho_2  -\alpha + 3\beta     \right) -\rho_1 , & \tau_2=0, \,\,   \qquad   &  \tau_4=1,   &   b_z = \frac{1}{2}\left( \rho_2  +\alpha -\beta     \right) -\rho_1 ,   
\end{align}
with $\rho_1 =\frac{1}{2}\left[ 2 \alpha ^3 (9 \beta +4)-\alpha ^2 (3 \beta  (9 \beta +10)+18 \gamma +8)+4 \alpha  (\beta  (9 \beta +27 \gamma +8)+12 \gamma +2) \right. $ \\
$ \left. -3 \alpha ^4 -3 (\beta +3 \gamma
)^2-4 \beta -8 \gamma -1 \right]^{1/2}$, $\rho_2= \alpha -\alpha ^2+3 \alpha  \beta  -3 \gamma$,
$a_x=a_y=a_z=1$.

\noindent This solution is evident rather involved. Easier solutions for the same scenario can be found by setting all coupling constants to zero. For instance, we find

\noindent {\bf Ta2:}\\
\begin{equation}
	g_1=g_2=g_3=0
\end{equation}
\begin{align}
	\mu_0 & = \frac{\omega_i^2 \omega_j^2}{a_x} , & \mu_2 & = \frac{\omega_i^2 + \omega_j^2}{a_x}, & \mu_4 & =\frac{1}{a_x}, & b_x & = a_x \omega_k^2, & i,j,k \in \text{Perm}(1,2,3), \\
	\nu_0 & =  \frac{\omega_l^2 \omega_m^2}{a_y}  , & \nu_2 & =  \frac{\omega_l^2 + \omega_m^2}{a_y} , & \nu_4 & = \frac{1}{a_y} , & b_y & = a_y \omega_n^2 &  l,m,n \in \text{Perm}(1,2,3),  \\
	\tau_0 & =  \frac{\omega_o^2 \omega_r^2}{a_z} , & \tau_2 & =  \frac{\omega_o^2 + \omega_r^2}{a_z} , & \tau_4 & = \frac{1}{a_z} , & b_z & = a_z \omega^2_s & o,r,,s \in \text{Perm}(1,2,3) ,
\end{align}

\noindent $k,n,s \in \text{Perm}(1,2,3).$ Notice that the corresponding coordinates in this case are those already used in \cite{pais1950field} for the case $N=3$ in there. In the next solution the first two equations in (\ref{equnmxy}) convert into the Pais-Uhlenbeck equation, whereas the last equation vanishes trivially.

\noindent {\bf Tb1:}\\
\begin{equation}
	g_1=\frac{\alpha  \tau _2-1}{2 \tau _2} , \,\,\,\,\, \quad
	g_2= -g_3-\tau _2  , \,\,\,\,\, \quad
	g_3 = \frac{-\tau _2^3\pm \sqrt{-2 \tau _2^2-2 \beta  \tau _2^4-\tau _2^6}}{2 \tau _2^2}  ,
\end{equation}
\begin{eqnarray}
	\mu_0 & =  &\beta +\tau _2 \left(g_3+\tau _2\right) ,  \quad   \mu_2=0,  \qquad  \mu_4=1,  \qquad      b_x = \frac{1+\alpha  \tau _2}{2 \tau _2} , \\
	\nu_0  & =  &\beta -g_3 \tau _2,   \qquad   \qquad \,\,\,\,    \nu_2=0,  \qquad    \nu_4 =1,  \qquad     b_y = \frac{1+\alpha  \tau _2}{2 \tau _2} ,  \\
	\tau_0 & = & 1,  \,\,\, \tau_2 = \frac{1\pm \sqrt{1+8 \alpha  (-\alpha  \beta +\gamma )}}{2 \alpha }  ,    \,\,\, \         \tau_4=0,   \,\,\, \   b_z = \tau _2 \left(\beta +2 g_3^2+2 g_3 \tau _2+\tau _2^2\right) ,   \qquad \,\,\,
\end{eqnarray}

\noindent Finally, we present a solution for which only the first equation in (\ref{equnmxy}) converts into the Pais-Uhlenbeck equation, whereas the other two vanish trivially.

\noindent {\bf Tc1:}\\
\begin{eqnarray}
	g_1&=&\frac{\gamma  \left[\gamma ^2-\left(\alpha  \gamma +\kappa _2\right) \mu _0+\beta  \mu _0^2\right] \nu _0-\kappa }{\gamma ^2 \kappa _1} , \,\,\,\,\, 
	g_2= \frac{\gamma  \left[\gamma ^2-\left(\alpha  \gamma +\kappa _2\right) \mu _0+\beta  \mu _0^2\right] \tau _0^2+\kappa  \nu _0}{\gamma ^2 \kappa _1 \tau _0} ,  \notag \\
	g_3&=& \frac{\gamma  \left[\kappa _1 \left(\kappa _2-\beta  \mu _0\right)-2 \mu _0 \left(\gamma ^2-\left(\alpha  \gamma +\kappa _2\right) \mu _0+\beta  \mu
		_0^2\right)\right] \nu _0 \tau _0^2+\kappa  \mu _0 \left(\tau _0^2-\nu _0^2\right)}{\gamma ^2 \kappa _1^2 \tau _0}
\end{eqnarray}
\begin{align}
	\mu_2&=\frac{\alpha  \mu _0+g_1 \nu _0+g_2 \tau _0-\gamma }{\mu _0},   \quad  &  \mu_4=1,  \quad  \,\, &    b_x = \alpha -\frac{\alpha  \mu _0+g_1 \nu _0+g_2 \tau _0-\gamma }{\mu _0} , \\
	\nu_2&=-g_1,  \,\,    \qquad   &  \nu_4 =0,  \quad  \,\, &    b_y = -\frac{g_1 \mu _0+g_3 \tau _0}{\nu _0}  ,  \,\, \,\, \quad   \qquad   \qquad \\
	\tau_2&=-g_2, \,\,   \qquad   &  \tau_4=0,    \quad  \,\,&   b_z = -\frac{g_2 \mu _0+g_3 \nu _0}{\tau _0} ,   
\end{align}
with $a_x=a_y=a_z=1$, $\kappa_1 =\nu _0^2+\tau _0^2$, $\kappa_2 =\mu _0^2+\nu _0^2+\tau _0^2$,\\
$\kappa = \gamma ^2 \tau _0^2 \left\{2 \alpha  \gamma ^3 \mu _0-\gamma ^4-\kappa _2 \left(\kappa _1-\beta  \mu _0+\mu _0^2\right){}^2+\alpha  \gamma  \left[\left(\beta
-\mu _0\right) \mu _0-\kappa _1\right] \left(\kappa _1+2 \mu _0^2\right) \right.$\\
$ \left. -\gamma ^2 \left[\alpha ^2 \mu _0^2-3 \kappa _1 \mu _0-2 \mu _0^3+\beta  \left(\kappa
_1+2 \mu _0^2\right)\right] \right\}$. In this transformation $\mu_0, \nu_0, \tau_0$ remain free parameters.

\subsection{Transformed Hamiltonians}
Next we carry out a Legendre transformation to obtain the corresponding Hamiltonians for the viable Lagrangians obtained in the previous section. With $p_x = a_x \dot{x}$, $p_y = a_y \dot{y}$, $p_z = a_z \dot{z}$ as momenta we compute
\begin{equation}
	{\cal H} (x,y,z,\dot{x}, \dot{y},\dot{z})  = p_x \dot{x} +p_y \dot{y} +p_z \dot{z} - {\cal L}(x,y,z,\dot{x}, \dot{y},\dot{z}) .   \label{Legtr}
\end{equation}
Using the expressions for the three Hamiltonians in (\ref{Ham1}), (\ref{Ham2}), (\ref{Ham3}) we identify
\begin{eqnarray}
	{\cal H}_{\text{Ta1}} &=&   \left(1-2 \alpha +3 \alpha ^2\right)     {\cal H}_1 +  \left( 3+\frac{\beta -3 \alpha  \beta }{\gamma }\right) {\cal H}_2   + \frac{3 \alpha -1}{\gamma } {\cal H}_3  ,   \\ 
	{\cal H}_{\text{Ta2}}&=& \left( \frac{\omega _k^4}{a_x}+\frac{\omega _n^4}{a_y}+\frac{\omega _s^4}{a_z}    \right) {\cal H}_1   - \frac{1}{\gamma}  \left( \frac{\omega_i^2 + \omega_j^2}{a_x} +\frac{\omega_l^2 + \omega_m^2}{a_y} + \frac{\omega_o^2 + \omega_r^2}{a_z} \right)  {\cal H}_2 \\ &&+ \frac{1}{\gamma} \left( \frac{\omega _k^2}{a_x}+\frac{\omega _n^2}{a_y}+\frac{\omega _s^2}{a_z}    \right)   {\cal H}_3  ,   \quad      \\
	{\cal H}_{\text{Tb1}} &=&     \left(  2 \alpha^2 + \tau_2^2  \right)   {\cal H}_1 + 2 \left( 1-\frac{\alpha  \beta }{\gamma } \right) {\cal H}_2   + \frac{2 \alpha}{\gamma}  {\cal H}_3  ,   \\ 
	{\cal H}_{\text{Tc1}}&=& \left[  \alpha ^2-\beta +\mu _0+\frac{\alpha  \left(\kappa _2-\beta  \mu _0\right)}{\gamma } \right] {\cal H}_1 + \left[ 1-\frac{\beta  \left(\alpha  \gamma +\kappa _2-\beta  \mu _0\right)}{\gamma ^2} \right]{\cal H}_2 \\
	&& +  \frac{\alpha  \gamma +\kappa _2-\beta  \mu _0}{\gamma ^2}  {\cal H}_3 .  \notag
\end{eqnarray}
Reading off the coefficients $c_{4,5,6}$ from these expressions, we can use (\ref{c1sol})-(\ref{c3sol}) to determine the constants $c_{1,2,3}$ and thus the corresponding Poisson tensors. 
\begin{figure}[h]
	\centering         
	\begin{minipage}[b]{1\textwidth}      
		\includegraphics[width=0.48\textwidth]{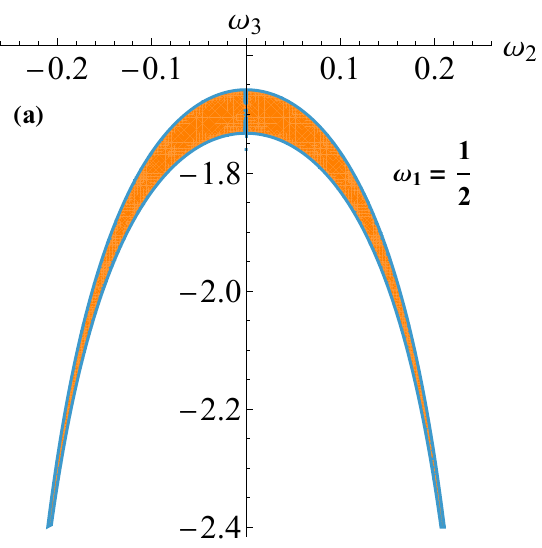}
		\includegraphics[width=0.48\textwidth]{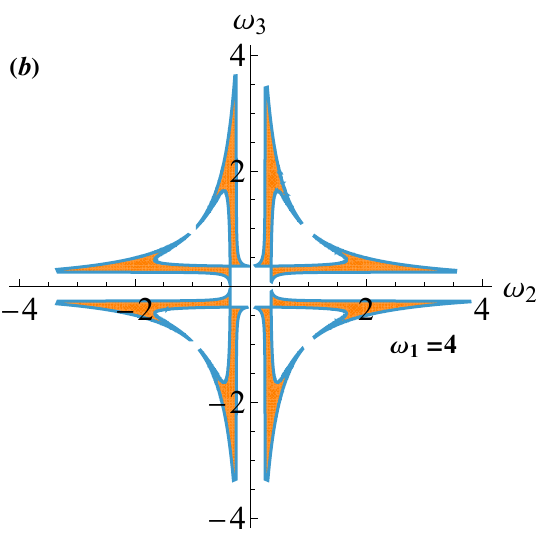}
	\end{minipage}   
	\caption{Sample regions in parameter space for which ${\cal H}_{\text{Ta1}}$ (panel a) and ${\cal H}_{\text{Tc1}}$ with $\mu_0=1$, $\kappa_2 = 2$ (panel b) are positive-definite.} 
	\label{figure1}
\end{figure}

The key question is which of these transformations yield positive-definite Hamiltonians, as can be verified by inspecting the signs of the coefficients in (\ref{e36}). It turns out that ${\cal H}_{\text{Ta2}}$ is always positive-definite when $a_x>0$, $a_y>0$, $a_z>0$, whereas ${\cal H}_{\text{Tb1}}$ is never positive for any choice of the $\omega_{1,2,3}$. The transformed Hamiltonians ${\cal H}_{\text{Ta1}}$ and ${\cal H}_{\text{Tc1}}$ are positive-definite only in a specific region in parameter space, as depicted for a sample set of frequencies in figure \ref{figure1}.

\section{Sixth-order PU model with interaction terms}

We finish with some remarks on the effects of adding a generic interaction term to the system, similarly as was investigated for the fourth order version of the PU model \cite{pais1950field}. We therefore consider systems with an added potential
\begin{equation}
	{\cal H}_i^{\text{int}} \left( q, \dot{q}, \ddot{q} , \dddot{q}, q_{4t},  q_{5t}  \right) = 	{\cal H}_i \left( q, \dot{q}, \ddot{q} , \dddot{q} q_{4t},  q_{5t}  \right) +  W(q),    \qquad i= 1,2,3,
\end{equation}	 	
the vector field will change to
\begin{align}
	\vec{v}_{\text{int}}  = \left(\dot{q},\ddot{q},q_{(3t)},q_{(4t)},q_{(5t)}, - \alpha q_{4t} - \beta \ddot{q} - \gamma q - W'\left( q \right) \right).
\end{align}
Verifying whether the Lie derivative (\ref{Lieder}) vanishes for this field with $J_{1,2,3}$ we find that $L_{	{V}_{\text{int}} } (J_1) = 0$, $L_{	{V}_{\text{int}} } (J_2) \neq 0$, $L_{	{V}_{\text{int}} } (J_3) \neq 0$, i.e. ${V}_{\text{int}} $ is only a Poisson vector field for $J_1$. Adding interaction terms of the form $W(\dot{q})$, $W(\ddot{q})$, $W(\dddot{q})$, $W(q_{4t})$, $W(q_{5t})$ leads to vector fields that are not Poissonian for any of the  $J_{1,2,3}$. Thus similarly as in the fourth order case, the multi-Hamiltonian structure ceases to exist in the presence of potentials.

\section{Conclusion}

We have shown that the sixth-order PU oscillator admits a full Tri-Hamiltonian formulation supported by a six-dimensional Abelian Lie algebra of dynamical symmetries. Within this framework, three mutually compatible Poisson tensors $J_{1,2,3}$ together with their associated Hamiltonians $H_{1,2,3}$ generate identical flows, and the Lie–derivative action iteratively builds an infinite hierarchy of conserved quantities in involution. This multi-Hamiltonian picture systematises the construction of ghost-free representatives and clarifies how symmetry, integrability, and stability intertwine for HTDT. In comparison to our previous analysis in \cite{FFT}, where a Bi-Hamiltonian structure was utilised for the fourth-order system, we require here the full Tri-Hamiltonian structure. This may suggest that for the PU system of order $2N$ one requires an $N$-Hamiltonian structure.

We identified conditions under which the models are ghost-free. In particular, while no single ${\cal H}_i$ is positive-definite, when expanded in the positive-definite building blocks $H_{ij}$, suitable combinations, such as $H= H_{12}+H_{23}+H_{13}$, yield positive-definite energies without altering the flow. The resulting positivity regions in parameter demand the full Tri-Hamiltonian data with non-vanishing $c_1,c_2,c_3$ and are encoded by simple frequency-ordered inequalities (\ref{HB1}), (\ref{HB2}), providing a practical recipe for selecting physically viable representatives. The Hamiltonians for the degenerate and partially degenerate cases cannot be expressed in terms of positive-definite quantities. This regime requires further investigations.  

Building on the structures above, we explicitly derived three-dimensional first-order representations that are dynamically equivalent to the PU model. This was achieved by mapping the sixth-order equation to families of coupled second-order systems and exhibited concrete parameter sets realizing one-to-one, two-to-one, and three-to-one embeddings, together with their Hamiltonians. These constructions expose a broad class of lower-order models that inherit the PU dynamics and, when the above positivity criteria are met, remain ghost-free. Finally, we assessed the effect of interactions and found that generic potentials spoil the multi-Hamiltonian structure—leaving only $J_1$ as a viable Poisson tensor—thus pinpointing the structural fragility of integrability and positivity under deformations.  Further investigations are needed to identify possible interaction terms (or deformations) that preserve at least a weakened multi-Hamiltonian structure. 

While our treatment is purely classical, we have provided numerous candidates that are likely to yield a consistent, ghost-free quantum theory. It is clear which classical Hamiltonians inevitably lead to non-physical quantum systems. Promising candidates include all positive-definite versions; however, uniqueness is lost, since all of them describe the same classical dynamics. An interesting problem for future investigation is to unravel the precise properties of these quantum theories, returning to an old question posed by Wigner \cite{wigner1950eq}: ``Do the equations of motion determine the quantum mechanical commutation relations?'' In principle, several outcomes are possible: the quantum versions may still coincide in some sense \cite{kaup1990quant}, they may be inequivalent \cite{scherer1993inequ,irac1993quantum}, or certain quantised versions may turn out to be inconsistent \cite{irac1993quantum}. General frameworks to address this problem have been developed \cite{carinena2000quantum,marmo2005alt,gutt1999equ}, that we will explore in future work \cite{FFTinprep}.

 \medskip

\noindent {\bf Acknowledgments}: AlF acknowledges funding from the Max Planck Society’s Lise Meitner Excellence Program 2.0. BT is supported by a City St George's, University of London Research Fellowship.

\newif\ifabfull\abfulltrue

\end{document}